\documentclass{epl}

\title{Topological Defects and the Spin Glass Phase of
Cuprates}
\shorttitle{Topological Defects $\dots$}
\author{N. Hasselmann\inst{1} \and A. H. Castro Neto\inst{2} \and 
C. Morais Smith\inst{3}}
\institute{
  \inst{1} Max-Planck-Inst. f. Physik komplexer Systeme, 
N\"othnitzer Str. 38, D-01187 Dresden \\
  \inst{2} Dept.~of Physics, Boston University, 
Boston MA 02215 \\
  \inst{3} Institut de Physique Th\'eorique, P\'erolles, CH-1700
Fribourg
}

\begin{document}

\maketitle

\begin{abstract}
We propose that the spin glass  phase of cuprates is 
due to the proliferation of topological defects of a spiral 
distortion of the antiferromagnet order. Our theory explains 
straightforwardly 
the simultaneous existence of short range incommensurate magnetic correlations 
and complete $a$-$b$ symmetry breaking
in this phase. We show via a renormalization group 
calculation that the collinear $O(3)/O(2)$ symmetry is unstable towards
the formation of local non-collinear correlations.
A critical disorder strength is identified beyond which topological defects
proliferate already at zero temperature.

\end{abstract}

\section{Introduction}
Until recently, the spin glass (SG) phase  
of $\rm La_{2-x}Sr_xCuO_4$ (LSCO), with $0.02<x<0.05$, was believed to have 
short ranged commensurate correlations, with incommensurate (IC)
correlations appearing only 
at the onset of superconductivity  $x\sim0.05$ \cite{yamada98}.
New data on LSCO have
however revealed IC spin correlations also in the SG regime 
\cite{wakimoto00a,wakimoto00b,matsuda00}.
The short range nature of the correlations may point to the existence of a
strongly disordered stripe glass, in accord with our previous analysis
\cite{hasselmann99}
which predicted an instability of the stripe array
towards a disorder dominated phase. 
Within the picture of disordered stripes,
one is however immediately confronted with a problem of two fundamentally
different scales: the inter-stripe distance, $\ell_s$, which scales as
$\ell_s\sim 1/x$ \cite{yamada98} and 
the average separation, $\ell_d$,
between disorder centers (Sr ions)
which scales as $\ell_d \sim 1/\sqrt{x}$. A conflict arises because
for pinned stripes, both scales are associated with the charge sector.
As $\ell_d\ll \ell_s$ in the SG
regime, the stripes must break up into short segments
or drastically reduce the on-stripe charge density to take advantage of
the disorder potential \cite{branko}. 

Such a conflict does not
arise in models
which assume that
holes localize individually 
near their dopants, as indicated by transport measurements 
\cite{keimer92},
and act as frustration centers to
the AF order \cite{aharony88,glazman90,gooding97,cherepanov99}. 
A polarization of the frustration centers can then lead to the
appearance of IC correlations \cite{shraiman89a}. In such a scenario,
the scale $\ell_s$ exists only in the spin sector whereas $\ell_d$ is
characteristic of the charge distribution. 
In the following we revisit this model, assuming that 
holes produce dipolar frustration, and
show that it can well explain the experiments. 
Our analysis sheds 
light on the origin of the cluster spin glass phase
which in our analysis arises from 
topological defects of a spiral spin texture.

Because dipolar frustration produces a long range distortion of the AF
order, this problem can be treated within a continuum field theory approach. 
A non-linear $\sigma$-model (NL$\sigma$M) for this problem reads
\cite{glazman90,cherepanov99}
\begin{equation}
\label{dipham}
\tilde{H}_{\rm col}
=\frac{\rho_s}{2T} \int d^2{\bf x} \left( \partial_\mu {\bf n} \right)^2
+\frac{\rho_s}{T}\int d^2{\bf x} \ {\bf f}_\mu \cdot \partial_\mu {\bf n}, 
\end{equation}
where $\rho_s$ is the spin stiffness, $T$ the temperature and
$\bf n$ a three component unit vector representing the local
staggered moment (we set $k_B=1$). 
The quenched field
${\bf f}_\mu$ represents the dipoles, with zero mean and a Gaussian 
distribution.
An analysis of (\ref{dipham})
leads to a finite correlation length at $T=0$ for any finite disorder strength
and to fair agreement with  the temperature and doping dependence of the
correlation length of LSCO \cite{glazman90,cherepanov99}.
As it stands, the model however predicts commensurate AF correlations, 
in contradiction with experiments
\cite{wakimoto00a,wakimoto00b,matsuda00}.

It is well known that the dipole model can lead to
IC correlations
\cite{shraiman89a}. The Hamiltonian (\ref{dipham}) favors
the formation of a spiral phase, with a
non-zero average twist $\partial_\mu {\bf n}$ of the AF order and a
simultaneous alignment of the dipoles, 
$\left< {\bf f}_\mu \right> \neq 0$, as long as
the lattice and spin degrees of freedom of dipoles are annealed.
While the spiral phase has been found to be unstable in the case of
an annealed hole distribution towards a local enhancement of the
spiral pitch, this instability arises from the fermionic susceptibility
\cite{auerbach}
an hence is absent in the model considered here, where all fermionic
degrees of freedom are quenched.
Further, as the lattice part of the ${\bf f}_\mu$ vector should reflect
the symmetries of the underlying lattice, a discrete set of
favored lattice vectors for the formation of the spiral exists. 
Thus, the $a$-$b$ (or  square lattice) symmetry breaking associated 
with the formation of spiral correlations can have truly long
range order. 
The continuous symmetry of spin space
on the other hand
inhibits long range magnetic order in the 2D system
for either finite temperature or finite disorder. The 
experimental observation of a macroscopic $a$-$b$ asymmetry 
\cite{matsuda00} but very short
spin correlation lengths thus clearly motivate the study of the dipole model.

We here investigate the dipole model assuming  
a random spatial distribution of the dipoles
but allowing for non-zero ordered moments. The non-collinearity
of the ground state which arises from ordered moments
requires a serious modification of previous collinear approaches for
two reasons: First, the symmetry breaking scheme of non-collinear magnets 
leads to the appearance of three rather than two Goldstone modes which 
couple to disorder. Second, a spiral phase allows for pointlike
topological defects (TD's) with $Z_2$ charges \cite{kawamura98}, 
which originate from a chiral degeneracy
of the spiral ground state and 
which are absent in collinear models.
Below, we investigate the influence of disorder on both
the Goldstone modes, using a renormalization group (RG) approach, and
on the creation of TD's, using a free energy argument.
A strongly disordered regime, as associated with
a SG, is only found once topological defects of
the spin texture are accounted for. The situation encountered here
thus resembles that of disordered planar XY
models, where the coupling of the disorder to spin waves
was also found to lead to a simple  
renormalization of the spin stiffness  whereas a phase transition
to a disordered phase is driven by a proliferation of TD's \cite{scheidl97}.
There are however profound differences between these 
simpler XY models and the present model.
Specifically, the renormalization of the spin
stiffness in the later case is primarily driven by the curvature
of the order parameter space while in the former case the renormalization
arises from the presence of TD's. Further,
the influence of TD's on the
renormalization of non-collinear Heisenberg models
is at present not understood and a unified renormalization group
approach which accounts for both TD's and curvature terms is 
not available \cite{kawamura98}. We therefore derive the RG
equations without incorporating TD's while we address the
importance of TD's in a separate analysis below,
where we determine the critical disorder threshold of the unbinding 
transition. Below this threshold, but not above, the RG 
decribes the system correctly.

\section{Derivation of the model}

As in presence of spiral correlations
the $O(3)$ symmetry of the spins is broken completely,
the $O(3)/O(2)$ AF model (\ref{dipham}) is not 
adequate. We need a formulation of the problem which incorporates
the order parameter for a spiral, which 
is an element of $O(3)\times O(2)/O(2)$ 
\cite{azaria93}. Thus, there are three instead of the two Goldstone
modes of collinear magnets. 
A possible representation of the local spiral order
is in terms of orthonormal ${\bf n}_k$, $k=1,2,3$, with
$n^a_k n^a_q=\delta_{kq}$.
A derivation of a continuum field theory for a spiral state
from a lattice Heisenberg model can be found in \cite{klee96}.
Using ${\bf S}_{ij}= {\bf n}_1 \cos ({\bf k}_S \cdot {\bf r}_{ij})
- {\bf n}_2 \sin ({\bf k}_S \cdot {\bf r}_{ij})$ and 
${\bf n}_3={\bf n}_1 \times{\bf n}_2$, where ${\bf S}_{ij}$ is the 
spin at the lattice site ($i,j$), ${\bf k}_S=(\pi,\pi)+{\bf q}_S$
and ${\bf q}_S$ is the IC ordering wave vector of the 
spiral, 
the effective classical Hamiltonian 
can be written in the form
\begin{equation}
\label{spiral1}
\tilde{H}=
\frac{1}{2} \int d^{2}{\bf x} \ p_{k} (\partial_{\mu} {\bf n}_{k})^{2}
+ s_\mu \int d^2{\bf x} \  {\bf n}_1 \cdot \partial_\mu {\bf n}_2
\end{equation}
where we ignore small anisotropies of order ${\bf q}_S^2$
in the stiffnesses $p_k$. Roughly, 
$p_{1,2}\simeq J/(2 T)$ and $p_3\simeq 0$. 
The vector $\bf s$ is to lowest order
given by ${\bf s}= J {\bf q}_S/T$. 
The second term
makes this Hamiltonian unstable, which simply shows
that the pure Heisenberg model does not support a
spiral ground state. We show below that this
term is
cancelled by a term originating from the coupling of
the spins to the dipoles.
Thus, the ordered dipoles stabilize the spiral
phase. 

We use a phenomenological form for the coupling of the spiral order
parameter to the dipoles.
The dipoles locally cant the spin order and couple
to the first spatial derivatives of the ${\bf n}_k$ fields.
To generate the frustration produced by the dipoles we thus
introduce a minimal coupling \cite{hertz78} in the first term
of (\ref{spiral1}), i.e. we replace $(\partial_\mu {\bf n}_k)^2$
with
$[ (\partial_\mu - i {\bf B}_\mu \cdot {\bf L} ){\bf n}_k]^2$
where ${\bf B}_\mu$ is a random dipole field. 
The components of $\bf L$ are $3\times 3$
matrix representations of angular momenta which generate rotations about
the three spin axes, with
$-i {\bf B}_\mu \cdot {\bf L} \  {\bf n}_k=
{\bf B}_\mu \times {\bf n}_k$.
Let us write 
${\bf B}_\mu = \left<{\bf B}_\mu \right>_D + {\bf Q}_\mu$
so that $\left<{\bf Q}_\mu \right>_D=0$, where $\left< \dots \right>_D$ 
is the disorder average. 
We obtain the following Hamiltonian for the spiral
in presence of disorder,
\begin{equation}
\label{spiral2}
\tilde{H}=
\frac{1}{2} \int d^{2}{\bf x}  p_{k} (\partial_{\mu} {\bf n}_{k})^{2}
+ \int d^{2} {\bf x} \ p_{k}  \partial_{\mu} {\bf n}_{k} \cdot {\bf Q}_{\mu}
\times {\bf n}_{k} ,
\end{equation}
where the ordered part of the dipole field
cancels the second term in (\ref{spiral1}). Thus,
$p_k \partial_\mu {\bf n}_k \cdot 
\left<{\bf B}_{\mu} \right>_D \times {\bf n}_k + 
s_\mu {\bf n}_1 \cdot \partial_\mu {\bf n}_2=0$. 
As ${\bf q}_S \propto {\bf s}$, this equation relates the incommensurability
linearly to the density of ordered dipoles.
The remaining part of the dipole field, $\bf{Q}_\mu$, is a quenched
variable with zero mean and we assume Gaussian short ranged statistics,
$\left<Q_{\mu}^{a}({\bf x}) Q_{\nu}^{b}({\bf y)}\right>_D=\lambda
 \delta({\bf x-y})\ \delta_{ab}\ \delta_{\mu \nu}$.
In absence 
of disorder, the Hamiltonian is $O(3)\times O(2)/O(2)$ symmetric.
The model reduces to (\ref{dipham}) in the case $p_{1,2}=0$ with
$p_3=\rho_s/T$, ${\bf n}_3={\bf n}$ and ${\bf f}_{\mu} =
{\bf Q}_{\mu} \times {\bf n}$. 
In Fourier space, the disorder coupling 
takes the form of a correlated random field coupling
$\int \frac{d^2{\bf q}}{(2 \pi)^2} \  
{\bf n}_k({-\bf q}) \cdot {\bf h}_k({\bf q})$,
with
\begin{equation}
\label{rfc}
{\bf h}_k({\bf q})= 
i p_{k} q_\mu \int d^2{\bf x}\ ({\bf Q}_\mu \times {\bf n}_k)
e^{i {\bf q }\cdot {\bf x}}
\end{equation}
and correlations $\left< {h}_k^a({\bf q}) 
{h}^{a^\prime}_{k^\prime}({\bf q}^\prime) \right>_D
\propto \delta({\bf q}-{\bf q}^\prime) |{\bf q}|^2$.
Using an Imry-Ma type argument it can be shown \cite{cherepanov99} that 
this coupling is marginal in two dimensions and that therefore a RG 
analysis is required to investigate the scaling of the disorder term.

\section{Renormalization}

\begin{figure}
\twofigures[width=5cm]{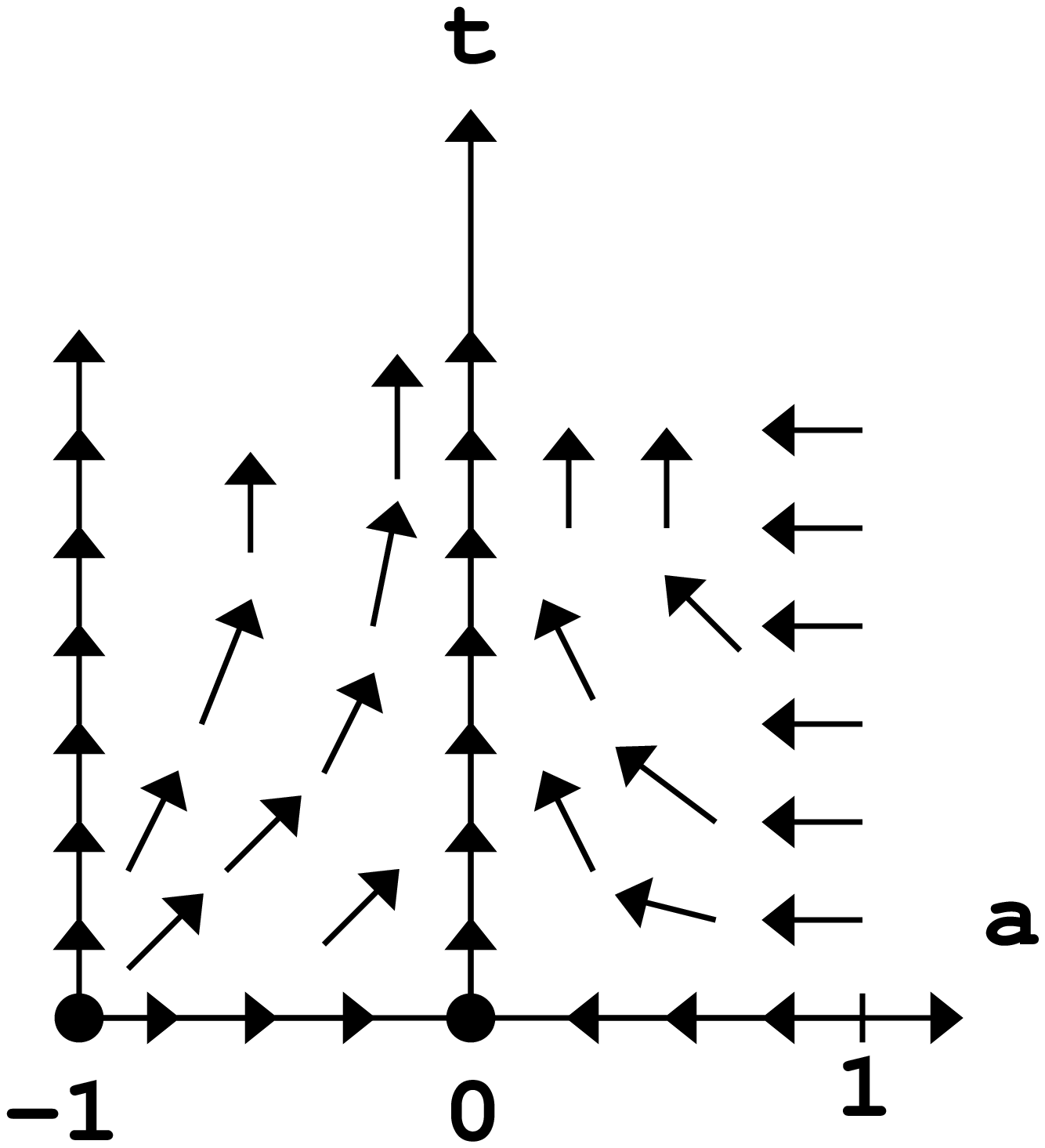}{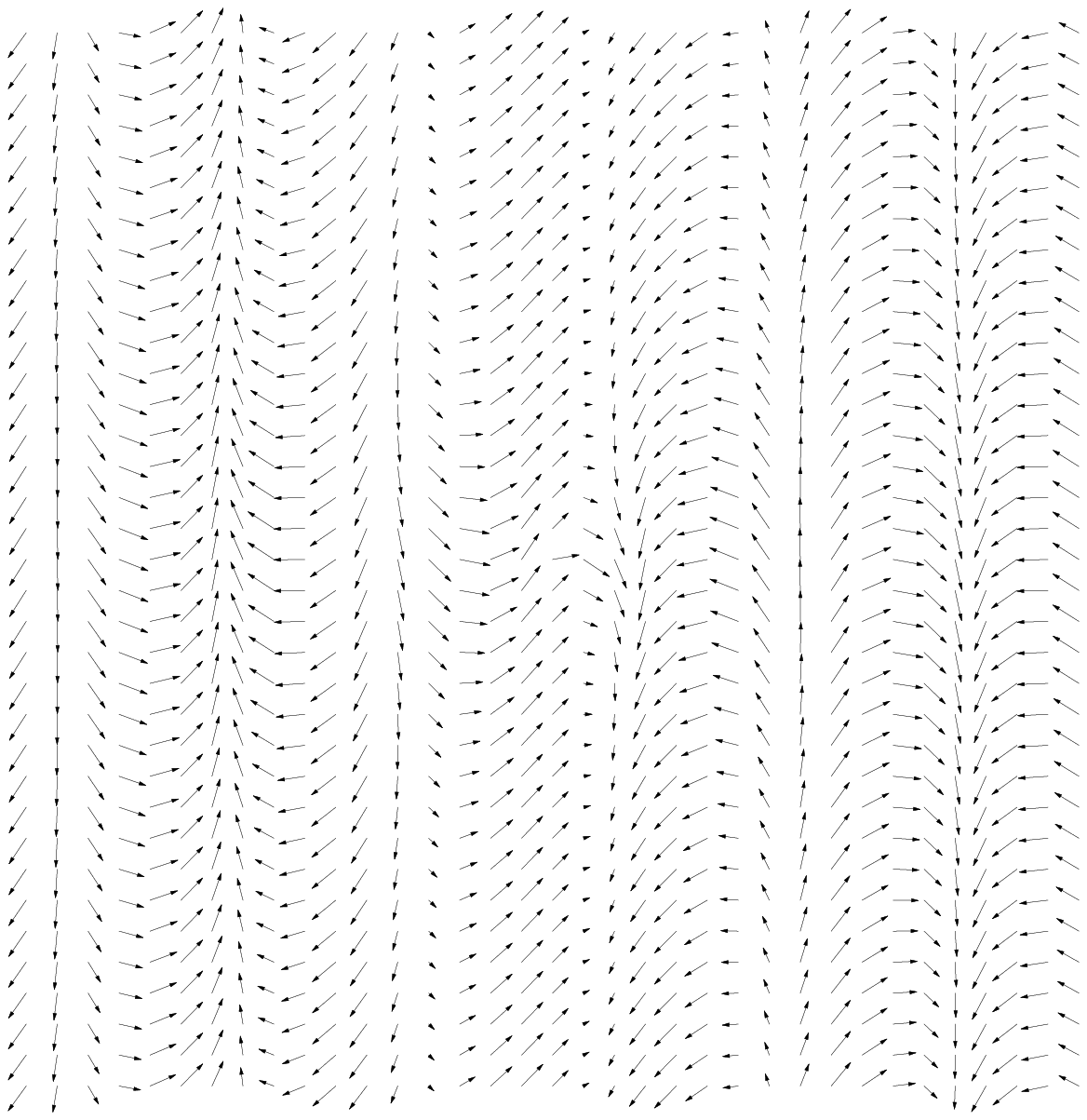}
\caption{RG flow for $\lambda\ll t$.}
\label{rgflow}
\caption{Single topological defect 
of a spiral with $a\ge0$. The arrows represent the local staggered
moments and the spin vector of ordered dipoles point out of the plane.}
\label{topopic}
\end{figure}

The RG is obtained in the
$SU(2)$ representation \cite{apel92} 
where $n_k^a =  \mbox{tr} \left[ \sigma^a  
g  \sigma^k g^{-1} \right]/2$
($\sigma^a$ are Pauli matrices and $g\in SU(2)$). Introducing
$A_\mu^a = -i \mbox{tr} \left[ \sigma^a  g^{-1}  
\partial_\mu g \right] /2$ \cite{polyakov75}, 
eq.~(\ref{spiral2}) can be written in the form,
\begin{equation} 
\label{finalstripe}
\tilde{H}=\frac{1}{t} \int d^{2}{\bf x} \left[ {\bf A}_\mu^2 + a {A_\mu^z}^2
\right] + 
2 \int d^{2}{\bf x} \ p_k \   \epsilon_{ijk} \ \epsilon_{abc} \
 A_\mu^i \ n_j^a \ n_k^c \ {Q}_\mu^b \ 
\end{equation}
where 
$t^{-1}=2(p_1+p_3)$ and $a=(p_1-p_3)/(p_1+p_3)$.
At the point $a=0$ 
the symmetry is enhanced to $O(3)\times O(3)/O(3) \simeq O(4)/O(3)$
while at $a=-1$ the model is collinear.
The one-loop RG equations are obtained by splitting the field $g$ 
into slow and fast modes,
$g=g^\prime \exp (i$ $\phi^a   \sigma^a )$ and tracing out the fast 
modes $\phi^a$ with fluctuations in the range $[\Lambda,1]$ (where the 
original UV cutoff is 1).
For $\lambda\ll t$ the RG equations are given by
\begin{eqnarray}
\label{RGT}
4 \pi \partial_\ell t & = & 2 (1-a) t^2
+ (2 -a + a^2) \lambda t \\ \nonumber
4 \pi \partial_\ell a  & = & - 4 a (1+a) t
- a (1+a)(3-a) \lambda
\end{eqnarray}
where $\ell = \ln(\Lambda)$.
For $\lambda=0$, these equations describe the RG of a clean spiral 
\cite{apel92}, while for the collinear point $a=-1$, the equations
reproduce the RG of the stiffness for disordered
collinear models \cite{cherepanov99}. From (\ref{RGT}) it is seen
that there are two fixed points for $a$, see also fig.~\ref{rgflow}.
 The collinear point $a=-1$ is 
unstable whereas $a=0$ is stable, 
irrespective of the disorder.   
The coupling
to weak disorder thus only renormalizes the spin stiffness but does not
not lead to any new fixed points.

The validity of
(\ref{RGT}) is however limited to high temperatures, for low temperatures
the RG of $\lambda$ must be accounted for.  
We  have  calculated the RG of $\lambda$ using the 
method described in \cite{cherepanov99}. However, for
$a\neq0$ new
disorder coupling terms appear.
Specifically, the RG generates transverse correlated fields
which couple to the ${\bf n}_{1,2}$ vectors even at $a=-1$ 
(i.~e.~$p_{1,2}=0$), whereas in
the original Hamiltonian with $a=-1$ they only couple to ${\bf n}_3$,
see eq.~(\ref{rfc}). 
Such transverse correlated fields
{\em destroy the collinear fixed point}.
Thus, even if the original AF order is collinear (i.e. in absence
of dipole ordering), the disorder
drives the system to a non-collinear state.
We recover the RG equations for the collinear model obtained
in \cite{cherepanov99} only if we ignore the correlated transverse
field coupling.
A purely
collinear analysis is thus not valid in presence of dipoles and cannot
describe the low temperature regime correctly. 
This is in fact expected,
because a random
canting of the spins destroys the remaining $O(2)$
spin symmetry of the collinear model.
We conjecture that the 
system will always flow to the point $a=0$, although this remains
to be proven. 

At the highest symmetry point $a=0$ however,  
no new coupling terms are generated. It is then straightforward to arrive
at the following RG equations, valid for
$a=0$ but any initial ratio of $\lambda/t$,
\begin{eqnarray}
2 \pi \partial_\ell t  = t^2
+ \lambda t & ; \ \ \ \ \ &
4 \pi \partial_\ell \lambda = \lambda^2
\end{eqnarray}
which can be simplified through $z=t+\lambda/2$ to get
$2 \pi \partial_{\ell} z = z^2$. 
So for $a=0$ the presence of
disorder leads to an additive renormalization of the stiffness,
$t\to t+\lambda/2$. 
In presence of any amount of disorder, 
the IC correlation length $\xi$ at $T=0$ is 
finite, as can be inferred from an integration of the RG equation,
yielding
$\xi \propto \exp(C \lambda_0^{-1})$ at $T=0$ with some constant $C$.
While the disorder scales to strong coupling,
the relative disorder strength with respect
to the stiffness, $\lambda/t$, always scales to zero so that at
long wavelengths the disorder becomes less relevant. This is surprisingly
different to the situation with $a=-1$ fixed \cite{cherepanov99},
where the ratio $\lambda/t$ was found to diverge below a certain
initial value of $\lambda_0/t_{0}$ which was interpreted as the scaling
towards a new disorder dominated regime. Thus, if one correctly takes into
account non-collinearity, this cross over to  a strongly 
disordered phase disappears. 

To summarize, within the perturbative RG analysis
disorder  leads only to a simple renormalization of the spin stiffness
which in turn leads to a finite correlation length already at $T=0$. 
As mentioned already in the introduction, disorder may
have a more dramatic effect via the creation of TD's in the spin 
texture. Below, we analyse the stability of the spin 
system against the creation of such defects.

\section{Topological defects}
TD's in spirals are related to the chiral degeneracy
of the spiral, i.e. a spiral can turn clock- or anti-clock wise 
\cite{wintel94,kawamura98}. These defects are pointlike and at
a TD the spiral changes its chirality. 
Saddle point solutions of TD's  
for spirals are of the form \cite{wintel94}
\begin{eqnarray}
\label{topog}
g({\bf x})=\exp[i m^a \sigma^a \Psi({\bf x})/2],
\end{eqnarray}
where $\bf m$ is a space independent unit vector and $\Psi({\bf x})$ satisfies
${\bf \nabla}^2 \Psi({\bf x})=0$ with singular solutions
$\Psi(x,y)=\arctan(y/x)$.
In fig. \ref{topopic} 
the spin distribution around an isolated defect in a
homogeneous spiral 
is shown for $a\ge0$.
The energy of a TD diverges logarithmically
with the linear system size $R$, 
$\beta E=[1+ (m^z)^2 a] \pi \ln R / 2 t$, 
so for $a<0$ the lowest energy defects have $m^z=\pm 1$ while for 
$a>0$, $m^z=0$ is preferred. 
Because of the logarithmic divergence of the energy, isolated
TD's are absent in
clean systems at sufficiently low enough temperatures and in this
regime the predictions of the NL$\sigma$M analysis should hold. Indeed,
NL$\sigma$M predictions for related frustrated Heisenberg models
were found to be in excellent agreement with
numerical simulations at low temperatures \cite{southern93}. 
Only at higher temperatures
an abrupt decrease of the correlation length was observed which 
has been attributed to the appearance of unpaired TD's
\cite{wintel95,southern93}. 
It has been argued, that the
high-temperature regime is similar to the equivalent phase
of XY models, as both are governed by the plasma phase of
the Coulomb gas model \cite{wintel95}.
The spiral defects differ however from their XY counterparts in
that the logarithmic divergence of
the energy appears only for length scales smaller than the correlation length
obtained from the NL$\sigma$M analysis. In an expansion
around the saddle points the defects do not decouple from the spin
waves even at the lowest (Gaussian) level. Also, TD's of spirals
carry a $Z_2$ charge, whereas XY vortices have $Z$ charges. Despite
these differences, the critical temperature, at which free TD's first
appear can in both cases be well estimated 
using  a free energy argument \cite{southern00}.  

A comparison to XY models is further usefull as the 
disorder coupling we employed 
is  very similar to the one appearing in XY models with randomly fluctuating
gauge fields \cite{scheidl97}. In these models, 
if one ignores vortices, 
the influence of the disorder was shown to
amount to a simple renormalization of the spin stiffness 
\cite{scheidl97},
and no disordering transition as a function of the disorder strength
is found. Once TD's are included in the
analysis, randomness leads
to a disordered phase even at $T=0$ through the creation of unpaired
defects if the disorder fluctuations
are stronger than some critical value \cite{scheidl97,nattermann95}. 
Similar to the estimate of the critical temperature at which thermal 
fluctuations lead
to the appearance of TD's , the 
critical disorder strength is well
estimated from the free energy of an isolated defect in presence of 
disorder \cite{scheidl97,cha95}. Here, we use this approach to calculate
the critical disorder strength for the spiral state.
The free energy of a defect state is given by
$\beta F=(1+ (m^z)^2 a)\pi \ln R/(2 t) - \left<\mbox{ln} Z_d\right>_D$,
where the second term contains the contribution of disorder,
\begin{eqnarray} \nonumber
Z_d=\int d^2{\bf y} \exp \left( -2 \int d^2{\bf x}\  p_k \ \epsilon_{ijk} 
\epsilon_{abc}\  A_\mu^i  n_j^a  n_k^c   Q_\mu^b \right)
\end{eqnarray}
with ${\bf A}_\mu$, ${\bf n}_k$ derived from (\ref{topog}).
Using the replica trick one finds at low
temperatures 
$\beta F=2 p_1 \pi (1-\sqrt{8 \lambda/\pi}) \ln R$ for $a\le0$ and
$\beta F=(p_1+p_3) \pi (1-\sqrt{8 \lambda/\pi}) \ln R$ for $a \ge 0$.
Thus, at low temperatures and $\lambda<\lambda_c=\pi/8$ free defects are absent
and the system is well described within the NL$\sigma$M analysis above.
For strong disorder,
$\lambda>\lambda_c$, the creation of
TD's is favorable even at $T=0$, leading
to a proliferation of free TD's and
much shorter correlation lengths than predicted from the
NL$\sigma$M, without affecting the incommensurability. 
Anticipating the analogy to XY models, for $\lambda>\lambda_c$
we expect an IC
correlation length roughly given by 
$\xi \propto \exp (b/\sqrt{\lambda-\lambda_c})$ (with some constant $b$)
\cite{scheidl97}. However,
instead of the divergence of $\xi$ at $\lambda=\lambda_c$ 
a crossover to the NL$\sigma$M result is expected.

\section{Comparison with experiment}
Experiments found the incommensurability to scale roughly linearly
with doping which points to a doping independent ordered fraction
of dipoles. 
To judge, whether or not TD's play a role in the
LSCO SG phase, we must estimate $\lambda$.  
We can use as a lower bound for $\lambda$ the result obtained
from the collinear analysis \cite{cherepanov99} where 
a disorder parameter equivalent to ours, but defined on
the much smaller scale of the AF unit cell, was used. From a fit
of the correlation lengths behavior at $x<0.02$ and large
temperatures, it is obtained $\lambda\simeq 20 x$. In this regime
of $x$, the low temperature phase has long range AF order and
a collinear analysis is well justified. 
This relation should remain valid also in the SG regime,
a view consistent with experiments where 
the width of the distribution of  internal magnetic
fields (i.e. local staggered moments) was found to
increase simply linearly with doping, with no detectable
change on crossing the AF/SG phase boundary \cite{niedermayer98}.
If we use
our above estimate for the critical disorder strength
we find remarkably a critical $x_c\sim 0.02$. 
Considering
that $\lambda\simeq 20 x$ is a conservative lower bound of $\lambda$ 
at the long length scales relevant to spirals,
we conclude that in the entire SG phase, free TD's
will be present already at $T=0$. 
Thus, 
we expect at $T=0$ IC correlation lengths which roughly
follow 
$\xi \sim \exp (b/\sqrt{20 x-\lambda_c})$. 

Our suggestion, that the IC correlations are
related to ordered dipolar frustration centers can be directly
tested experimentally in co-doped samples $\rm La_{2-x}Sr_xCu_{1-z}Zn_zO_4$. 
It was argued \cite{korenblit99} that co-doping has
the effect of renormalizing the density of dipoles to lower
values  $x\rightarrow x (1-\gamma  z)$ (experiments indicate $\gamma \sim 2$ 
\cite{korenblit99}) 
as Zn atoms placed close to
a localized hole strongly distort the hole wavefunction which 
leads to a reduction or complete destruction of the AF frustration induced
by the hole. 
Thus, co-doping with
Zn has two effects: First, it lowers the amount of frustration
in the sample and thus enhances the correlation length 
\cite{korenblit99,huecker99}. Secondly,
Zn doping also lowers the total amount of ordered dipoles which
leads to a decrease of the incommensurability by a factor $1-\gamma z$
which should be observable experimentally. 

\section{Conclusions} We propose a novel description of the SG
phase as a strongly disordered spiral state, which can account for 
the IC correlations and the $a$-$b$ asymmetry without invoking 
any kind of charge order.
We show that a collinear analysis  
is inadequate to describe the strongly disordered regime.
No sharp transition towards a 
disorder dominated phase is found within the NL$\sigma$M model analysis
and TD's of the spiral must be accounted for to explain
the extremely short correlations observed in experiments \cite{curro99}.
The picture of the cluster SG phase as proposed in \cite{gooding97} 
has some similarities to our work 
in that it invokes the presence of defects.
However, in this picture, the defects are those of
an $O(3)/O(2)$ spin system (skyrmions). Also, while 
the skyrmion model predicts IC correlations
only for $x>0.05$ \cite{gooding97}, our analysis includes
IC correlations from the outset. 

\acknowledgments
We thank R.~M.~Noack, N.~J.~Curro, P.~C.~Hammel and A.~R.~Bishop 
for stimulating
discussions.
N.~H. acknowledges 
the hospitality
of the  CNLS
at LANL and the
Inst.~Phys.~Th\'eorique, Univ.~Fribourg,
where part of this work was completed. A.~H.~C.~N. acknowledges support from 
a LANL CULAR grant. C.~M.~S. was supported by Swiss National Foundation
under grant 620-62868.00.

\end{document}